\documentclass[superscriptaddress,prl,twocolumn,amssymb,nofootinbib,aps]{revtex4-2}

\usepackage{braket}
\usepackage{color}
\usepackage{graphicx}
\usepackage{dcolumn}
\usepackage{bm}

\begin{document}

\title{GHz configurable photon pair generation from a silicon  nonlinear interferometer}

\author{Jonathan Frazer}
\affiliation{Quantum Engineering Technology Labs, H. H. Wills Physics Laboratory and Department of Electrical \& Electronic Engineering, University
of Bristol, BS8 1FD, United Kingdom}
\affiliation{Quantum Engineering Centre for Doctoral Training, H. H. Wills Physics Laboratory and Department of Electrical and Electronic
Engineering, University of Bristol, Tyndall Avenue, BS8 1FD, United Kingdom}
\author{Takafumi Ono}
\affiliation{Faculty of Engineering and Design, Kagawa University, 1-1 Saiwaicho, Takamatsu, Kagawa 760-0016, Japan
}
\author{Jonathan C. F. Matthews}
\affiliation{Quantum Engineering Technology Labs, H. H. Wills Physics Laboratory and Department of Electrical \& Electronic Engineering, University
of Bristol, BS8 1FD, United Kingdom}
\date{\today}

\begin{abstract}
\noindent Low loss and high speed processing of photons is central to architectures for photonic quantum information. 
High speed switching enables non-deterministic photon sources and logic gates to be made deterministic, while the  speed with which quantum light sources can be turned on and off impacts the clock rate of photonic computers and the data rate of quantum communication. Here we use lossy carrier depletion modulators in a silicon waveguide nonlinear interferometer to modulate photon pair generation at 1~GHz without exposing the generated photons to the phase dependent parasitic loss of the modulators. The super sensitivity of nonlinear interferometers reduces power consumption compared to modulating the driving laser. This can be a building block component for high speed programmabile, generalised nonlinear waveguide networks.
\end{abstract}

\maketitle
        
Networks of linear interferometers form universal linear optics~\cite{reck1994experimental} for a range of photonic quantum information applications~\cite{carolan2015universal}. Silicon's optical nonlinearity enable photon pair sources~\cite{Fukuda2005Jun} that have been integrated at the input of programmable photonic networks~\cite{Bao2023Apr}, and integrating reconfigurable nonlineartity throughout optical networks would further enhance capability and reconfigurability~\cite{Krenn2017}. Nonlinear intereferometers (NLI) provide reconfigurable nonlinearity. They were first described~\cite{Yurke1986Jun} for phase sensitivity below the standard quantum limit and have since been explored in many contexts including imaging~\cite{Lemos2014Aug}, generalised nonlinear networks~\cite{Paterova2020May},  photon source engineering~\cite{Gu2019Mar, cui2020quantum}, tests of nonlocality~\cite{Krenn2021Mar}, mid-infrared (mid-IR) greenhouse gas sensing without mid-IR detectors~\cite{Lidner21}, in silicon photonics~\cite{Ono2019Mar}, and in graph quantum photonics~\cite{Bao2023Apr}.

Silicon-on-insulator (SOI) photonics is an attractive platform for quantum technology~\cite{silverstone2016silicon} due to its high refractive index contrast enabling high component density, a strong $\chi_3$ non-linearity, monolithic integration with electronics~\cite{TaskerFrazer23} and for scalable manufacture. Key components realised include high quality photon sources~\cite{Paesani19NCOM} and low propagation loss $\sim 100$~kHz bandwidth thermo-optic phase shifters~(TPS)~\cite{Harris2014May}. 
However, because TPS rely on local heating, thermal crosstalk results in unwanted phase shifts in nearby waveguides~\cite{Jacques2019Apr} and the total heat dissipation of larger photonic circuits limits integration with cryogenically cooled superconducting single photon detectors~\cite{Chakraborty2020Oct}. 
        
Simultaneously fast and low-loss phase modulators remain an outstanding challenge in SOI quantum photonics. They are needed to multiplex non-deterministic photon sources to increase photon generation rates~\cite{Kaneda15}, for feedforward~\cite{prevedel2007high} and adaptive state processing~\cite{wang2017experimental}. High speed modulation is implemented in silicon photonics with carrier depletion modulators (CDM)~\cite{Reed10} that use the plasma dispersion effect. Free carriers made available by doping are swept from the waveguide with a biasing electric field. This yields changes to both real and imaginary parts of the refractive index and hence parasitic loss in conjunction with the phase shift~\cite{Reed2005Jan}. The presence of free carriers in the waveguide core and cladding also contribute to excess loss with typical values of $1 - 8$~dB/mm. This is  problematic for quantum technologies because optical loss is catastrophic for key quantum effects such as squeezing and entanglement. However the benefit of CDMs are their speed --- tens of gigahertz are acheived in classical applications~\cite{Reed2014Aug} and as such they have been used in high speed quantum key distribution~\cite{Sibson2017Feb}.
        
Here we present a GHz-reconfigurable photon pair source that uses the high bandwidth of a CDM in conjunction with a non-linear interferometer to avoid the deleterious loss of CDMs. We use gated detection from superconducting nanowire single photon detectors (SNSPDs) to verify modulation speed. The super-sensitive nonlinear interference effect intrinsically reduces power consumption by a factor of four compared to modulating a classical laser pump on chip with a CDM. This method  avoids the need for direct modulation on generated photons, thereby guarding delicate quantum state generation from loss induced by high-speed modulators.
        
A nonlinear interferometer consisting of two identical photon pair sources in series generates the state,
\begin{equation}
    \ket{\psi} = P(1+\exp^{i \phi_p}) \ket{1}_s \ket{1}_i 
\end{equation}
where $\phi_{p}$ is the relative phase between pump beams in each source, $P$ is the pump power and subscripts on kets label signal and idler modes. We assume $P$ is set sufficiently low that higher order Fock terms are negligible at measurement. Nonlinear interference between four-wave mixing sources~\cite{Ono2019Mar} outputs a coincidence detection described by the fringe,
\begin{equation}
    \label{eq:fringe_vis}
    f(\phi_{p}) = \frac{1}{2} (1 + v(R)\cos(n \phi_{p} + \Phi_0))
\end{equation}
where $n=2$, and $\Phi_0$ captures all other phase terms. $v(R) = 2R/(1 + R^2)$ represents the visibility in terms of the ratio $R = P_{2}/P_{1}$ of the probabilities of pair production in each source. The fringe expected from a Mach-Zehnder interferometer is of the same form, but $n=1$. This difference in fringe frequency is from the SFWM phasematching condition for degenerate pump photons,
\begin{equation}
    \phi = 2 \phi_p  + \phi_i + \phi_s .
\end{equation}
Therefore, the signature of nonlinear interference between SFWM sources is the doubling of fringe period, compared to a classical linear interferometer fringe.
        
\begin{figure}
    \centering
    \includegraphics[width = \linewidth]{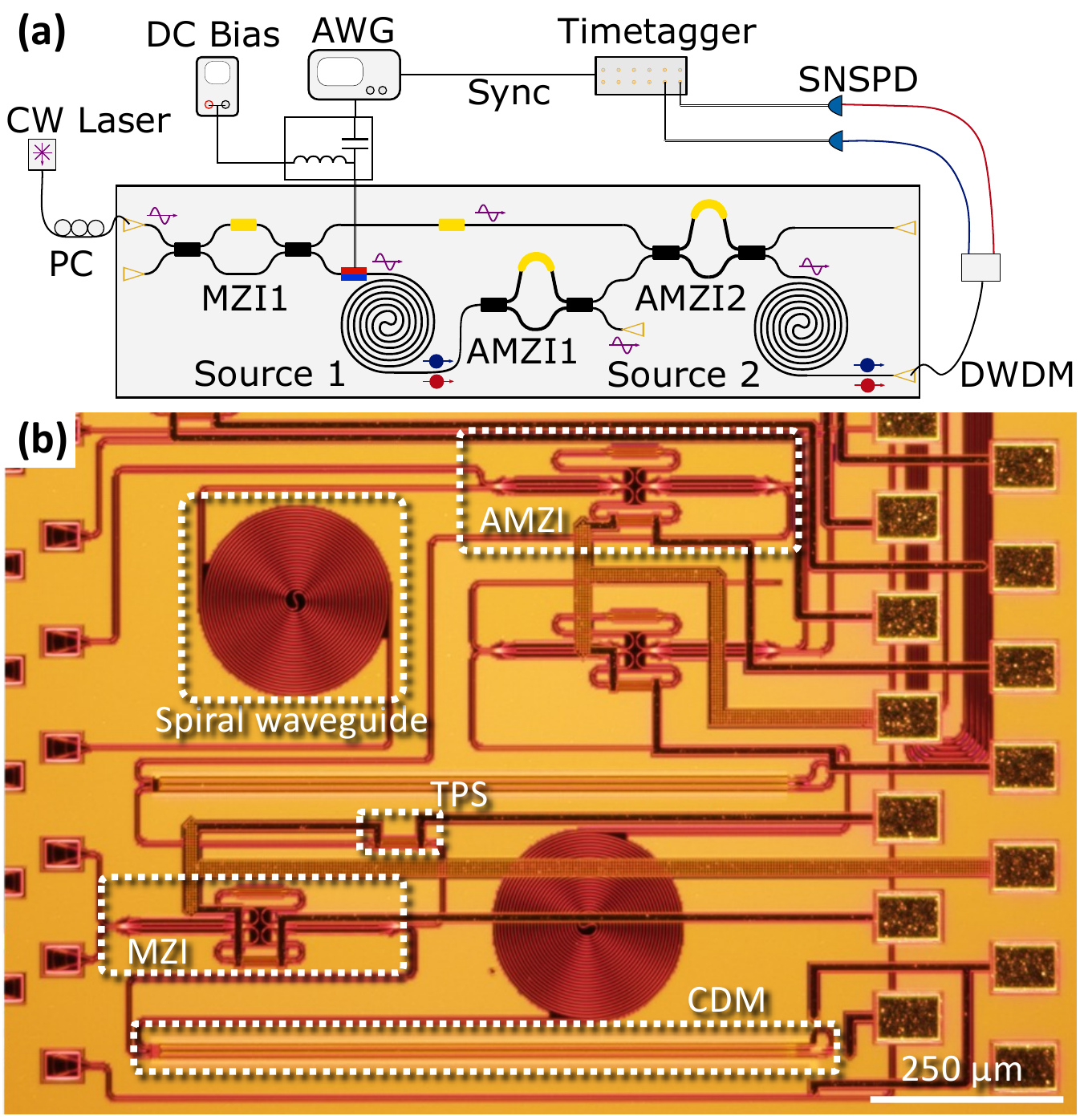}
    \caption{
        The GHz non-linear interference experiment. (a)~Schematic of the experiment, as described in the main text. 
        TPS (gold) are used to reconfigure the circuit and offset the CDM phase to minimise the modulation depth required for nonlinear interference. The CDM is depicted by the red/blue box in the lower arm of the interferometer. The reconfigurability of AMZI1 and AMZI2 allows for comparison between nonlinear interference between sources and classical interference in a Mach-Zehnder. (b) Microscope image the photonic integrated circuit used in the experiment.
    }
    \label{fig:chip_diagram}
\end{figure}
        
The photonic integrated circuit used is shown in Fig.~\ref{fig:chip_diagram}. It is of in-house design and fabricated by commercially outsourcing to IMEC using their ISIPP50G silicon photonic platform \cite{Rahim2019May}. The photonic circuit is mounted on an in-house designed printed circuit board (PCB) that carries signals for TPS and the CDM. Temperature stability is maintained using a Peltier element and PID controller (Arroyo 5240).
        
To operate the chip as a nonlinear interferometer, a continuous wave (CW) laser (Rio Orion), tuned to 1544.61~nm and amplified with an  erbium doped-fiber amplifier (Pritel), is coupled to the photonic chip via a fibre polarisation controller (PC), a v-groove fibre array and grating coupler. On chip, the pump is split into two paths with a Mach-Zehnder interferometer (MZI1), tuned to 50:50 beamsplitting to balance power between sources. Asymmetric Mach-Zehnder interferometers with path length difference $\Delta L = 90 \mu m$ are used as add-drop filters with a free-spectral range of $\sim6.4$~nm to discard the pump after first source (AMZI-1), and multiplex photon pairs into the second source (AMZI-2). Each photon pair source is a $\sim$1.1~cm spiral of single waveguide~\cite{Sharping2006Dec}. Photons are coupled off-chip via a grating coupler and separated into signal and idler channels using fibre based dense-wavelength-division multiplexers (DWDM). Superconducting nanowire single photon detectors (SNSPD,  Photonspot, $\sim$85\% system efficiency) detect the photons. All TPS are driven by commercial multichannel voltage sources (Qontrol~Q8iv). The CDM is driven with signals from an arbitrary waveform generator (AWG, Tektronix 70001A), amplified through an RF voltage voltage amplifier (SHF 810). A bias tee (SHF BT45) is used to combine the RF signal and a DC bias from a benchtop power supply. A TPS in the upper depicted path (without the CDM) is used to minimise coincidences when no RF signal is applied to the CDM.
        
\begin{figure}
    \centering
    \includegraphics[width=0.9\linewidth]{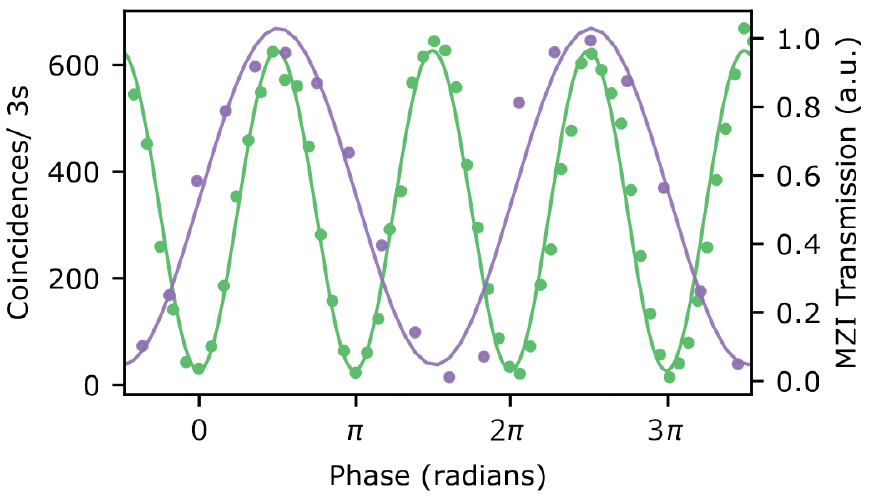}
    \caption{
        Linear and nonlinear interference with TPS. Purple data and line of fit from Eq.~\ref{eq:fringe_vis} ($n=1$) correspond to classical characterisation on right axis. Quantum interference data is plotted in green, with a line of fit from Eq.~\ref{eq:fringe_vis} ($n=2$). Classical and background subtracted quantum interference display visibilities of $99\%$ and $(96\pm1)\%$ respectively.
        }
    \label{fig:thermal_data}
\end{figure}

We verified non-linear interference occurs by comparison with a classical interference fringe on the same device. The chip is configured for classical interference by setting AMZI-1 to perform the identity operation and AMZI-2 to act as a 50:50 beamsplitter. For simplicity, the classical interference is measured by monitoring pump power at the device output. Fig~\ref{fig:thermal_data} shows raw and fitted data for classical linear interference and for coincidences output from the NLI configuration, in both cases scanning only the TPS in the upper arm. 
        
We demonstrated quantum interference modulated by the CDM. With the chip in its NLI configuration, the CDM is driven with a square wave generated by AWG at $10$~MHz and $1$~GHz. The driving signal and CDM are impedance matched with a $50~\Omega$ termination (Smiths Interconnect) wirebonded directly to the photonic chip. The heat dissipated by this termination prevents direct characterisation of the modulator $V_{\pi}$ due to thermal crosstalk. We therefore apply a small RF signal (300~mV at 100~kHz) to the modulator while the chip is configured for classical interference and extrapolate $V_{\pi}$ assuming a linear voltage-phase relationship. This provides a small-signal estimate of $V_{\pi} = 7.99 \pm 0.02$~V.

\begin{figure}
    \centering
    \includegraphics[width=\linewidth]{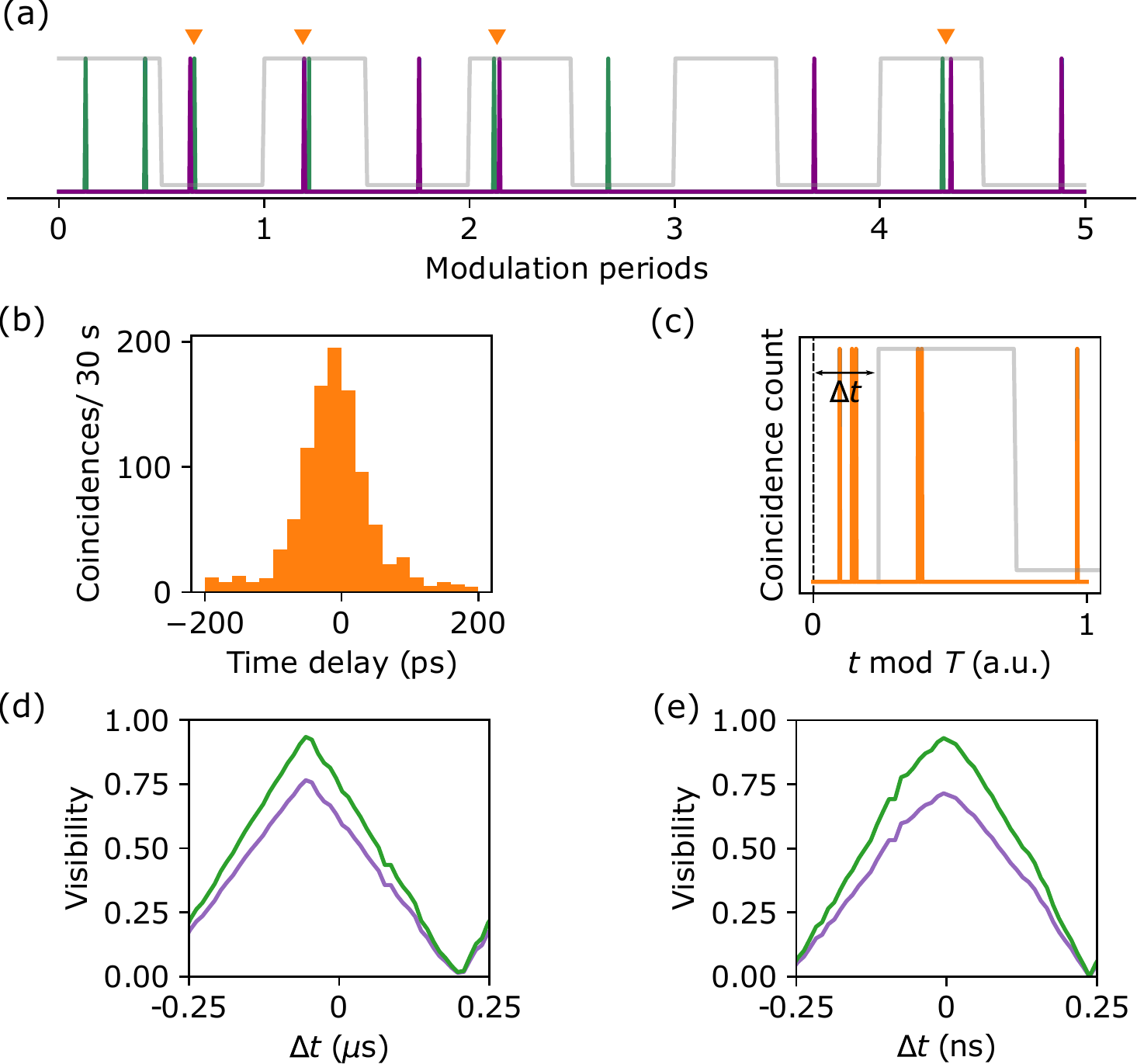}
    \caption{
        Coincidence counting of photon pairs. (a) Simulated data illustrating photon coincidences identified between 2 detection channels. Signal and idler pairs (green and purple) are recorded as a coincidences (red) when the delay between channels is less than a predetermined threshold. A square wave driving  signal of period $T$ that modulates coincidence rates is overlaid in grey. (b) Coincidences are recorded as a histogram showing the delay between channels with a sign given by the order of timetag events. The absolute timing reference of each timetag is typically discarded. (c) Coincidences may also be located in time by taking the midpoint of timetags from corresponding channels. This data can then be processed modulo $T$ (up to some constant delay) to correlate with the periodic signal used to drive the CDM. (d, e) Coincidence visibility with 10~MHz~(d) and 1~GHz~(e) modulation as the offset $\Delta t$ is applied to the time series of coincidence data, divided modulo the waveform period (see main text). Raw data is purple. Background-subtracted data  is green. Maximum visibility corresponds to data displayed in Fig.~\ref{fig:mod_coins}.
        }
        \label{fig:analysis}
\end{figure}

We observe coincidence rates from the device of $\sim$~100 counts per second. The probability of observing a coincidence within any given interval is given by $\langle n \rangle e^{-n}$, where $\langle n \rangle$ is the average coincidence rate in the same interval. Therefore, the probability of observing a coincidence within any given cycle of the modulation is very small, precluding a direct measurement of the interference with GHz driving signals. We therefore attempt to correlate the driving signal with the distribution of coincidences in time by analysing timetags modulo the period of the driving signal. Fig.~\ref{fig:analysis} depicts this analysis. 
        
\begin{figure}[h!]
    \centering
    \includegraphics[width = 0.8\linewidth]{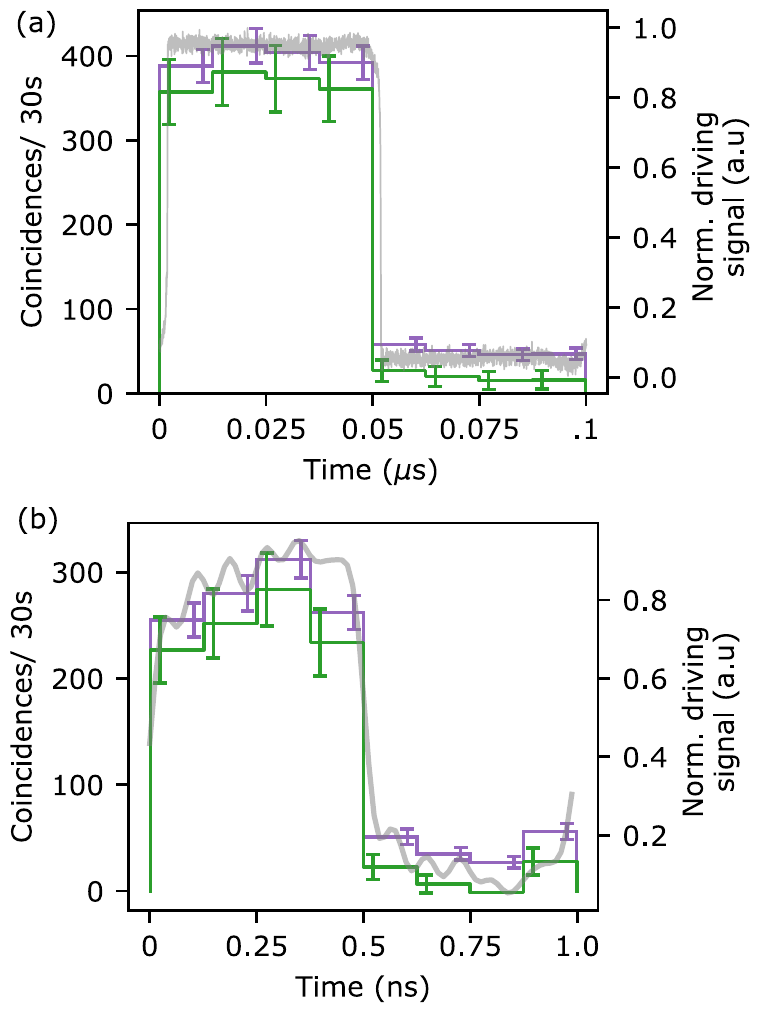}
    \caption{
        Raw (purple) and background subtracted (green) coincidence counts from 10~MHz (a) and 1~GHz (b) square waves applied to the modulator (left axes). Timetag data is collected for 30~s while the modulation is applied, before postprocessing to identify coincidences. The peak to peak voltage of the modulator has been set at 3.95~V and 4.0~V, respectively, with a fixed DC bias of 2.1~V. The normalised driving waveform is overlaid in grey for comparison (right axes). Errors are calculated assuming Poissonian statistics.
        }
    \label{fig:mod_coins}
\end{figure}

We synchronise the SNSPD timing logic (Swabian Instruments) with the AWG using a common $10$~MHz clock. We note that while the AWG and timetagger clock are synchronised, there is an unknown offset of up to $\frac{1}{2\Omega}$ between the driving signal and the coincidence data, where $\Omega$ is the driving frequency. To account for this we offset the photon time tags over a range of $\Delta t = \pm \frac{1}{4 \Omega} = \pm \frac{1}{4} T$, which is half the period of the drive waveform. We expect a square wave modulation in the coincidence data, and therefore as this offset is swept we expect the convolution of two square wave, which is a triangle wave peaked at the maximum coincidence visibility. Fig.~\ref{fig:mod_coins} shows the correlation between the modulation signal and the coincidence data where we have normalised the timetags to zero offset. We compute the interference visibility by summing coincidence counts from the high and low modulator states. We observe (Fig.~\ref{fig:mod_coins}) a raw visibility of $(78 \pm 1)\%$ and $(74 \pm 2)\%$ at $10$~MHz and $1$~GHz. We estimate the accidental coincidence rates by summing counts outside the histogram peak and dividing by the number of bins. This gives corrected visibilities of $(90 \pm 1)\%$ and $(89 \pm 1)\%$. The single photon counts also vary with $\phi_p$, although the fringe visibility is limited by detector dark counts, any distinguishing information between sources such as loss and other photon generation not contained within spiral waveguide arms of the NLI~\cite{Ono2019Mar}. The maximum observed single photon fringe visibility of $4.41 \pm 0.07 \%$ implies a total loss $\approx-13.5~$dB. We measure 6.5~dB loss in each spiral, and attribute the remaining  $\sim0.5$~dB loss to waveguide and AMZI transmission.
        
We attribute the limited coincidence visibility to a number of causes. The low-frequency, small-signal value of the voltage for $\pi$-phase, $V_{\pi}$, only gives an estimate as the phase shifting efficiency rolls off with frequency due to impedance mismatch with the modulator electrodes, or velocity mismatch to the optical signal. The values we used were determined empirically by halving $V_{\pi}$ and then scanning in steps of $50$~mV to maximise fringe visibility. A more accurate estimate of $V_{\pi}/2$ would improve fringe visibility. Additionally, thermal cross-talk from the modulator termination reduces fringe visibility by changing the heat distribution across the chip, despite the global temperature set by the Peltier element and temperature controller. Improved heat sinking or recalibration of TPS while the CDM is operated could mitigate this effect.

In this work, we have demonstrated high speed modulation of nonlinear interferometry in silicon photonics to modulate photon pair generation at GHz speed. This  does not expose the photon pairs generated to any parasitic loss of the CDM, and 
could allow for control over generation of other delicate quantum resources, without incurring insertion losses from high speed modulators. Because of the super-sensitive interference frequency of NLIs, the method enhances efficiency and reduces heat dissipation compared to modulation on the pump laser with a CDM at the input of the SOI chip. With the operation of silicon quantum photonics in the mid-IR~\cite{Rosenfeld20}, we anticipate on-chip NLIs can be combined with degenerate four wave mixing
and mid-IR spectroscopy to detect greenhouse gas species on chip without needing mid-IR detectors~\cite{Lidner21}. 
Fast modulation of the interference effect could then be used to move the gas detection signal to a side band away from electronic noise sources. We also anticipate fast integrated NLIs to be a building block of more general nonlinear interferometric networks with increased capability for quantum applications over universal linear optics, such as scaling up large NLI superlattices~\cite{Paterova2020May}, multiplexed sources~\cite{rudolph2018optical} and quantum state engineering~\cite{Gu2019Mar}.

\textbf{Acknowledgements and Contributions.} We thank J. F. Tasker, A. Laing and Giacomo Ferranti for useful discussions. The authors are grateful for technical assistance from L. Kling. The experiment was conceived by T.O. and J.C.F.M. The PIC was designed by T.O. J.F. performed the experiment and data analysis, supervised by J.C.F.M. All authors contributed to writing the manuscript. J.F. acknowledges support from EPSRC Quantum Engineering Centre for Doctoral Training EP/L015730/1 and an EPSRC iCASE Thales studentship. J.C.F.M. acknowledges support from European Research Council starting grant ERC-2018-STG 803665 and a Philip Leverhulme Prize.


\begin{thebibliography}{30}%
\makeatletter
\providecommand \@ifxundefined [1]{%
 \@ifx{#1\undefined}
}%
\providecommand \@ifnum [1]{%
 \ifnum #1\expandafter \@firstoftwo
 \else \expandafter \@secondoftwo
 \fi
}%
\providecommand \@ifx [1]{%
 \ifx #1\expandafter \@firstoftwo
 \else \expandafter \@secondoftwo
 \fi
}%
\providecommand \natexlab [1]{#1}%
\providecommand \enquote  [1]{``#1''}%
\providecommand \bibnamefont  [1]{#1}%
\providecommand \bibfnamefont [1]{#1}%
\providecommand \citenamefont [1]{#1}%
\providecommand \href@noop [0]{\@secondoftwo}%
\providecommand \href [0]{\begingroup \@sanitize@url \@href}%
\providecommand \@href[1]{\@@startlink{#1}\@@href}%
\providecommand \@@href[1]{\endgroup#1\@@endlink}%
\providecommand \@sanitize@url [0]{\catcode `\\12\catcode `\$12\catcode
  `\&12\catcode `\#12\catcode `\^12\catcode `\_12\catcode `\%12\relax}%
\providecommand \@@startlink[1]{}%
\providecommand \@@endlink[0]{}%
\providecommand \url  [0]{\begingroup\@sanitize@url \@url }%
\providecommand \@url [1]{\endgroup\@href {#1}{\urlprefix }}%
\providecommand \urlprefix  [0]{URL }%
\providecommand \Eprint [0]{\href }%
\providecommand \doibase [0]{https://doi.org/}%
\providecommand \selectlanguage [0]{\@gobble}%
\providecommand \bibinfo  [0]{\@secondoftwo}%
\providecommand \bibfield  [0]{\@secondoftwo}%
\providecommand \translation [1]{[#1]}%
\providecommand \BibitemOpen [0]{}%
\providecommand \bibitemStop [0]{}%
\providecommand \bibitemNoStop [0]{.\EOS\space}%
\providecommand \EOS [0]{\spacefactor3000\relax}%
\providecommand \BibitemShut  [1]{\csname bibitem#1\endcsname}%
\let\auto@bib@innerbib\@empty
\bibitem [{\citenamefont {Reck}\ \emph {et~al.}(1994)\citenamefont {Reck},
  \citenamefont {Zeilinger}, \citenamefont {Bernstein},\ and\ \citenamefont
  {Bertani}}]{reck1994experimental}%
  \BibitemOpen
  \bibfield  {author} {\bibinfo {author} {\bibfnamefont {M.}~\bibnamefont
  {Reck}}, \bibinfo {author} {\bibfnamefont {A.}~\bibnamefont {Zeilinger}},
  \bibinfo {author} {\bibfnamefont {H.~J.}\ \bibnamefont {Bernstein}},\ and\
  \bibinfo {author} {\bibfnamefont {P.}~\bibnamefont {Bertani}},\ }\bibfield
  {title} {\bibinfo {title} {Experimental realization of any discrete unitary
  operator},\ }\href@noop {} {\bibfield  {journal} {\bibinfo  {journal}
  {Physical review letters}\ }\textbf {\bibinfo {volume} {73}},\ \bibinfo
  {pages} {58} (\bibinfo {year} {1994})}\BibitemShut {NoStop}%
\bibitem [{\citenamefont {Carolan}\ \emph {et~al.}(2015)\citenamefont
  {Carolan}, \citenamefont {Harrold}, \citenamefont {Sparrow}, \citenamefont
  {Mart{\'\i}n-L{\'o}pez}, \citenamefont {Russell}, \citenamefont
  {Silverstone}, \citenamefont {Shadbolt}, \citenamefont {Matsuda},
  \citenamefont {Oguma}, \citenamefont {Itoh} \emph
  {et~al.}}]{carolan2015universal}%
  \BibitemOpen
  \bibfield  {author} {\bibinfo {author} {\bibfnamefont {J.}~\bibnamefont
  {Carolan}}, \bibinfo {author} {\bibfnamefont {C.}~\bibnamefont {Harrold}},
  \bibinfo {author} {\bibfnamefont {C.}~\bibnamefont {Sparrow}}, \bibinfo
  {author} {\bibfnamefont {E.}~\bibnamefont {Mart{\'\i}n-L{\'o}pez}}, \bibinfo
  {author} {\bibfnamefont {N.~J.}\ \bibnamefont {Russell}}, \bibinfo {author}
  {\bibfnamefont {J.~W.}\ \bibnamefont {Silverstone}}, \bibinfo {author}
  {\bibfnamefont {P.~J.}\ \bibnamefont {Shadbolt}}, \bibinfo {author}
  {\bibfnamefont {N.}~\bibnamefont {Matsuda}}, \bibinfo {author} {\bibfnamefont
  {M.}~\bibnamefont {Oguma}}, \bibinfo {author} {\bibfnamefont
  {M.}~\bibnamefont {Itoh}}, \emph {et~al.},\ }\bibfield  {title} {\bibinfo
  {title} {Universal linear optics},\ }\href@noop {} {\bibfield  {journal}
  {\bibinfo  {journal} {Science}\ }\textbf {\bibinfo {volume} {349}},\ \bibinfo
  {pages} {711} (\bibinfo {year} {2015})}\BibitemShut {NoStop}%
\bibitem [{\citenamefont {Fukuda}\ \emph {et~al.}(2005)\citenamefont {Fukuda},
  \citenamefont {Yamada}, \citenamefont {Shoji}, \citenamefont {Takahashi},
  \citenamefont {Tsuchizawa}, \citenamefont {Watanabe}, \citenamefont
  {Takahashi},\ and\ \citenamefont {Itabashi}}]{Fukuda2005Jun}%
  \BibitemOpen
  \bibfield  {author} {\bibinfo {author} {\bibfnamefont {H.}~\bibnamefont
  {Fukuda}}, \bibinfo {author} {\bibfnamefont {K.}~\bibnamefont {Yamada}},
  \bibinfo {author} {\bibfnamefont {T.}~\bibnamefont {Shoji}}, \bibinfo
  {author} {\bibfnamefont {M.}~\bibnamefont {Takahashi}}, \bibinfo {author}
  {\bibfnamefont {T.}~\bibnamefont {Tsuchizawa}}, \bibinfo {author}
  {\bibfnamefont {T.}~\bibnamefont {Watanabe}}, \bibinfo {author}
  {\bibfnamefont {J.-i.}\ \bibnamefont {Takahashi}},\ and\ \bibinfo {author}
  {\bibfnamefont {S.-i.}\ \bibnamefont {Itabashi}},\ }\bibfield  {title}
  {\bibinfo {title} {{Four-wave mixing in silicon wire waveguides}},\ }\href
  {https://doi.org/10.1364/OPEX.13.004629} {\bibfield  {journal} {\bibinfo
  {journal} {Opt. Express}\ }\textbf {\bibinfo {volume} {13}},\ \bibinfo
  {pages} {4629} (\bibinfo {year} {2005})}\BibitemShut {NoStop}%
\bibitem [{\citenamefont {Bao}\ \emph {et~al.}(2023)\citenamefont {Bao},
  \citenamefont {Fu}, \citenamefont {Pramanik}, \citenamefont {Mao},
  \citenamefont {Chi}, \citenamefont {Cao}, \citenamefont {Zhai}, \citenamefont
  {Mao}, \citenamefont {Dai}, \citenamefont {Chen}, \citenamefont {Jia},
  \citenamefont {Zhao}, \citenamefont {Zheng}, \citenamefont {Tang},
  \citenamefont {Li}, \citenamefont {Luo}, \citenamefont {Wang}, \citenamefont
  {Yang}, \citenamefont {Peng}, \citenamefont {Liu}, \citenamefont {Dai},
  \citenamefont {He}, \citenamefont {Muthali}, \citenamefont {Oxenl{\o}we},
  \citenamefont {Vigliar}, \citenamefont {Paesani}, \citenamefont {Hou},
  \citenamefont {Santagati}, \citenamefont {Silverstone}, \citenamefont
  {Laing}, \citenamefont {Thompson}, \citenamefont {O{'}Brien}, \citenamefont
  {Ding}, \citenamefont {Gong},\ and\ \citenamefont {Wang}}]{Bao2023Apr}%
  \BibitemOpen
  \bibfield  {author} {\bibinfo {author} {\bibfnamefont {J.}~\bibnamefont
  {Bao}}, \bibinfo {author} {\bibfnamefont {Z.}~\bibnamefont {Fu}}, \bibinfo
  {author} {\bibfnamefont {T.}~\bibnamefont {Pramanik}}, \bibinfo {author}
  {\bibfnamefont {J.}~\bibnamefont {Mao}}, \bibinfo {author} {\bibfnamefont
  {Y.}~\bibnamefont {Chi}}, \bibinfo {author} {\bibfnamefont {Y.}~\bibnamefont
  {Cao}}, \bibinfo {author} {\bibfnamefont {C.}~\bibnamefont {Zhai}}, \bibinfo
  {author} {\bibfnamefont {Y.}~\bibnamefont {Mao}}, \bibinfo {author}
  {\bibfnamefont {T.}~\bibnamefont {Dai}}, \bibinfo {author} {\bibfnamefont
  {X.}~\bibnamefont {Chen}}, \bibinfo {author} {\bibfnamefont {X.}~\bibnamefont
  {Jia}}, \bibinfo {author} {\bibfnamefont {L.}~\bibnamefont {Zhao}}, \bibinfo
  {author} {\bibfnamefont {Y.}~\bibnamefont {Zheng}}, \bibinfo {author}
  {\bibfnamefont {B.}~\bibnamefont {Tang}}, \bibinfo {author} {\bibfnamefont
  {Z.}~\bibnamefont {Li}}, \bibinfo {author} {\bibfnamefont {J.}~\bibnamefont
  {Luo}}, \bibinfo {author} {\bibfnamefont {W.}~\bibnamefont {Wang}}, \bibinfo
  {author} {\bibfnamefont {Y.}~\bibnamefont {Yang}}, \bibinfo {author}
  {\bibfnamefont {Y.}~\bibnamefont {Peng}}, \bibinfo {author} {\bibfnamefont
  {D.}~\bibnamefont {Liu}}, \bibinfo {author} {\bibfnamefont {D.}~\bibnamefont
  {Dai}}, \bibinfo {author} {\bibfnamefont {Q.}~\bibnamefont {He}}, \bibinfo
  {author} {\bibfnamefont {A.~L.}\ \bibnamefont {Muthali}}, \bibinfo {author}
  {\bibfnamefont {L.~K.}\ \bibnamefont {Oxenl{\o}we}}, \bibinfo {author}
  {\bibfnamefont {C.}~\bibnamefont {Vigliar}}, \bibinfo {author} {\bibfnamefont
  {S.}~\bibnamefont {Paesani}}, \bibinfo {author} {\bibfnamefont
  {H.}~\bibnamefont {Hou}}, \bibinfo {author} {\bibfnamefont {R.}~\bibnamefont
  {Santagati}}, \bibinfo {author} {\bibfnamefont {J.~W.}\ \bibnamefont
  {Silverstone}}, \bibinfo {author} {\bibfnamefont {A.}~\bibnamefont {Laing}},
  \bibinfo {author} {\bibfnamefont {M.~G.}\ \bibnamefont {Thompson}}, \bibinfo
  {author} {\bibfnamefont {J.~L.}\ \bibnamefont {O{'}Brien}}, \bibinfo {author}
  {\bibfnamefont {Y.}~\bibnamefont {Ding}}, \bibinfo {author} {\bibfnamefont
  {Q.}~\bibnamefont {Gong}},\ and\ \bibinfo {author} {\bibfnamefont
  {J.}~\bibnamefont {Wang}},\ }\bibfield  {title} {\bibinfo {title}
  {{Very-large-scale integrated quantum graph photonics}},\ }\href
  {https://doi.org/10.1038/s41566-023-01187-z} {\bibfield  {journal} {\bibinfo
  {journal} {Nat. Photonics}\ ,\ \bibinfo {pages} {1}} (\bibinfo {year}
  {2023})}\BibitemShut {NoStop}%
\bibitem [{\citenamefont {Krenn}\ \emph {et~al.}(2017)\citenamefont {Krenn},
  \citenamefont {Gu},\ and\ \citenamefont {Zeilinger}}]{Krenn2017}%
  \BibitemOpen
  \bibfield  {author} {\bibinfo {author} {\bibfnamefont {M.}~\bibnamefont
  {Krenn}}, \bibinfo {author} {\bibfnamefont {X.}~\bibnamefont {Gu}},\ and\
  \bibinfo {author} {\bibfnamefont {A.}~\bibnamefont {Zeilinger}},\ }\bibfield
  {title} {\bibinfo {title} {{Quantum Experiments and Graphs: Multiparty States
  as Coherent Superpositions of Perfect Matchings}},\ }\href@noop {} {\bibfield
   {journal} {\bibinfo  {journal} {Phys. Rev. Lett.}\ }\textbf {\bibinfo
  {volume} {119}},\ \bibinfo {pages} {240403} (\bibinfo {year}
  {2017})}\BibitemShut {NoStop}%
\bibitem [{\citenamefont {Yurke}\ \emph {et~al.}(1986)\citenamefont {Yurke},
  \citenamefont {McCall},\ and\ \citenamefont {Klauder}}]{Yurke1986Jun}%
  \BibitemOpen
  \bibfield  {author} {\bibinfo {author} {\bibfnamefont {B.}~\bibnamefont
  {Yurke}}, \bibinfo {author} {\bibfnamefont {S.~L.}\ \bibnamefont {McCall}},\
  and\ \bibinfo {author} {\bibfnamefont {J.~R.}\ \bibnamefont {Klauder}},\
  }\bibfield  {title} {\bibinfo {title} {{SU(2) and SU(1,1) interferometers}},\
  }\href {https://doi.org/10.1103/PhysRevA.33.4033} {\bibfield  {journal}
  {\bibinfo  {journal} {Phys. Rev. A}\ }\textbf {\bibinfo {volume} {33}},\
  \bibinfo {pages} {4033} (\bibinfo {year} {1986})}\BibitemShut {NoStop}%
\bibitem [{\citenamefont {Lemos}\ \emph {et~al.}(2014)\citenamefont {Lemos},
  \citenamefont {Borish}, \citenamefont {Cole}, \citenamefont {Ramelow},
  \citenamefont {Lapkiewicz},\ and\ \citenamefont {Zeilinger}}]{Lemos2014Aug}%
  \BibitemOpen
  \bibfield  {author} {\bibinfo {author} {\bibfnamefont {G.~B.}\ \bibnamefont
  {Lemos}}, \bibinfo {author} {\bibfnamefont {V.}~\bibnamefont {Borish}},
  \bibinfo {author} {\bibfnamefont {G.~D.}\ \bibnamefont {Cole}}, \bibinfo
  {author} {\bibfnamefont {S.}~\bibnamefont {Ramelow}}, \bibinfo {author}
  {\bibfnamefont {R.}~\bibnamefont {Lapkiewicz}},\ and\ \bibinfo {author}
  {\bibfnamefont {A.}~\bibnamefont {Zeilinger}},\ }\bibfield  {title} {\bibinfo
  {title} {{Quantum imaging with undetected photons}},\ }\href
  {https://doi.org/10.1038/nature13586} {\bibfield  {journal} {\bibinfo
  {journal} {Nature}\ }\textbf {\bibinfo {volume} {512}},\ \bibinfo {pages}
  {409} (\bibinfo {year} {2014})}\BibitemShut {NoStop}%
\bibitem [{\citenamefont {Paterova}\ and\ \citenamefont
  {Krivitsky}(2020)}]{Paterova2020May}%
  \BibitemOpen
  \bibfield  {author} {\bibinfo {author} {\bibfnamefont {A.~V.}\ \bibnamefont
  {Paterova}}\ and\ \bibinfo {author} {\bibfnamefont {L.~A.}\ \bibnamefont
  {Krivitsky}},\ }\bibfield  {title} {\bibinfo {title} {{Nonlinear interference
  in crystal superlattices - Light: Science {\&} Applications}},\ }\href
  {https://doi.org/10.1038/s41377-020-0320-1} {\bibfield  {journal} {\bibinfo
  {journal} {Light Sci. Appl.}\ }\textbf {\bibinfo {volume} {9}},\ \bibinfo
  {pages} {1} (\bibinfo {year} {2020})}\BibitemShut {NoStop}%
\bibitem [{\citenamefont {Gu}\ \emph {et~al.}(2019)\citenamefont {Gu},
  \citenamefont {Erhard}, \citenamefont {Zeilinger},\ and\ \citenamefont
  {Krenn}}]{Gu2019Mar}%
  \BibitemOpen
  \bibfield  {author} {\bibinfo {author} {\bibfnamefont {X.}~\bibnamefont
  {Gu}}, \bibinfo {author} {\bibfnamefont {M.}~\bibnamefont {Erhard}}, \bibinfo
  {author} {\bibfnamefont {A.}~\bibnamefont {Zeilinger}},\ and\ \bibinfo
  {author} {\bibfnamefont {M.}~\bibnamefont {Krenn}},\ }\bibfield  {title}
  {\bibinfo {title} {{Quantum experiments and graphs II: Quantum interference,
  computation, and state generation}},\ }\href
  {https://doi.org/10.1073/pnas.1815884116} {\bibfield  {journal} {\bibinfo
  {journal} {Proc. Natl. Acad. Sci. U.S.A.}\ }\textbf {\bibinfo {volume}
  {116}},\ \bibinfo {pages} {4147} (\bibinfo {year} {2019})}\BibitemShut
  {NoStop}%
\bibitem [{\citenamefont {Cui}\ \emph {et~al.}(2020)\citenamefont {Cui},
  \citenamefont {Su}, \citenamefont {Li}, \citenamefont {Liu}, \citenamefont
  {Li},\ and\ \citenamefont {Ou}}]{cui2020quantum}%
  \BibitemOpen
  \bibfield  {author} {\bibinfo {author} {\bibfnamefont {L.}~\bibnamefont
  {Cui}}, \bibinfo {author} {\bibfnamefont {J.}~\bibnamefont {Su}}, \bibinfo
  {author} {\bibfnamefont {J.}~\bibnamefont {Li}}, \bibinfo {author}
  {\bibfnamefont {Y.}~\bibnamefont {Liu}}, \bibinfo {author} {\bibfnamefont
  {X.}~\bibnamefont {Li}},\ and\ \bibinfo {author} {\bibfnamefont
  {Z.}~\bibnamefont {Ou}},\ }\bibfield  {title} {\bibinfo {title} {Quantum
  state engineering by nonlinear quantum interference},\ }\href@noop {}
  {\bibfield  {journal} {\bibinfo  {journal} {Physical Review A}\ }\textbf
  {\bibinfo {volume} {102}},\ \bibinfo {pages} {033718} (\bibinfo {year}
  {2020})}\BibitemShut {NoStop}%
\bibitem [{\citenamefont {Feng}\ \emph {et~al.}(2023)\citenamefont {Feng},
  \citenamefont {Zhang}, \citenamefont {Liu}, \citenamefont {Cheng},
  \citenamefont {Guo}, \citenamefont {Dai}, \citenamefont {Guo}, \citenamefont
  {Krenn},\ and\ \citenamefont {Ren}}]{Krenn2021Mar}%
  \BibitemOpen
  \bibfield  {author} {\bibinfo {author} {\bibfnamefont {L.-T.~F.}\
  \bibnamefont {Feng}}, \bibinfo {author} {\bibfnamefont {M.}~\bibnamefont
  {Zhang}}, \bibinfo {author} {\bibfnamefont {D.}~\bibnamefont {Liu}}, \bibinfo
  {author} {\bibfnamefont {Y.-J.}\ \bibnamefont {Cheng}}, \bibinfo {author}
  {\bibfnamefont {G.-P.}\ \bibnamefont {Guo}}, \bibinfo {author} {\bibfnamefont
  {D.-X.}\ \bibnamefont {Dai}}, \bibinfo {author} {\bibfnamefont {G.-C.}\
  \bibnamefont {Guo}}, \bibinfo {author} {\bibfnamefont {M.}~\bibnamefont
  {Krenn}},\ and\ \bibinfo {author} {\bibfnamefont {X.-F.}\ \bibnamefont
  {Ren}},\ }\bibfield  {title} {\bibinfo {title} {{On-chip quantum interference
  between the origins of a multi-photon state}},\ }\href
  {https://doi.org/10.1364/OPTICA.474750} {\bibfield  {journal} {\bibinfo
  {journal} {Optica}\ }\textbf {\bibinfo {volume} {10}},\ \bibinfo {pages}
  {105} (\bibinfo {year} {2023})}\BibitemShut {NoStop}%
\bibitem [{\citenamefont {Lidner}\ \emph {et~al.}(2021)\citenamefont {Lidner},
  \citenamefont {Herr}, \citenamefont {Wolf}, \citenamefont {Kie{\ss}ling},\
  and\ \citenamefont {K{\"{u}}nemann}}]{Lidner21}%
  \BibitemOpen
  \bibfield  {author} {\bibinfo {author} {\bibfnamefont {J.}~\bibnamefont
  {Lidner}, \bibfnamefont {Chiara amd~Lunz}}, \bibinfo {author} {\bibfnamefont
  {S.~J.}\ \bibnamefont {Herr}}, \bibinfo {author} {\bibfnamefont
  {S.}~\bibnamefont {Wolf}}, \bibinfo {author} {\bibfnamefont {J.}~\bibnamefont
  {Kie{\ss}ling}},\ and\ \bibinfo {author} {\bibfnamefont {F.}~\bibnamefont
  {K{\"{u}}nemann}},\ }\bibfield  {title} {\bibinfo {title} {{Nonlinear
  interferometer for Fourier-transform mid-infrared gas spectroscopy using
  near-infrared detection}},\ }\href@noop {} {\bibfield  {journal} {\bibinfo
  {journal} {Optics Express}\ }\textbf {\bibinfo {volume} {29}},\ \bibinfo
  {pages} {4035} (\bibinfo {year} {2021})}\BibitemShut {NoStop}%
\bibitem [{\citenamefont {Ono}\ \emph {et~al.}(2019)\citenamefont {Ono},
  \citenamefont {Sinclair}, \citenamefont {Bonneau}, \citenamefont {Thompson},
  \citenamefont {Matthews},\ and\ \citenamefont {Rarity}}]{Ono2019Mar}%
  \BibitemOpen
  \bibfield  {author} {\bibinfo {author} {\bibfnamefont {T.}~\bibnamefont
  {Ono}}, \bibinfo {author} {\bibfnamefont {G.~F.}\ \bibnamefont {Sinclair}},
  \bibinfo {author} {\bibfnamefont {D.}~\bibnamefont {Bonneau}}, \bibinfo
  {author} {\bibfnamefont {M.~G.}\ \bibnamefont {Thompson}}, \bibinfo {author}
  {\bibfnamefont {J.~C.~F.}\ \bibnamefont {Matthews}},\ and\ \bibinfo {author}
  {\bibfnamefont {J.~G.}\ \bibnamefont {Rarity}},\ }\bibfield  {title}
  {\bibinfo {title} {{Observation of nonlinear interference on a silicon
  photonic chip}},\ }\href {https://doi.org/10.1364/OL.44.001277} {\bibfield
  {journal} {\bibinfo  {journal} {Opt. Lett.}\ }\textbf {\bibinfo {volume}
  {44}},\ \bibinfo {pages} {1277} (\bibinfo {year} {2019})}\BibitemShut
  {NoStop}%
\bibitem [{\citenamefont {Silverstone}\ \emph {et~al.}(2016)\citenamefont
  {Silverstone}, \citenamefont {Wang}, \citenamefont {Bonneau}, \citenamefont
  {Sibson}, \citenamefont {Santagati}, \citenamefont {Erven}, \citenamefont
  {O'Brien},\ and\ \citenamefont {Thompson}}]{silverstone2016silicon}%
  \BibitemOpen
  \bibfield  {author} {\bibinfo {author} {\bibfnamefont {J.~W.}\ \bibnamefont
  {Silverstone}}, \bibinfo {author} {\bibfnamefont {J.}~\bibnamefont {Wang}},
  \bibinfo {author} {\bibfnamefont {D.}~\bibnamefont {Bonneau}}, \bibinfo
  {author} {\bibfnamefont {P.}~\bibnamefont {Sibson}}, \bibinfo {author}
  {\bibfnamefont {R.}~\bibnamefont {Santagati}}, \bibinfo {author}
  {\bibfnamefont {C.}~\bibnamefont {Erven}}, \bibinfo {author} {\bibfnamefont
  {J.}~\bibnamefont {O'Brien}},\ and\ \bibinfo {author} {\bibfnamefont
  {M.}~\bibnamefont {Thompson}},\ }\bibfield  {title} {\bibinfo {title}
  {Silicon quantum photonics},\ }in\ \href@noop {} {\emph {\bibinfo {booktitle}
  {2016 International Conference on Optical MEMS and Nanophotonics (OMN)}}}\
  (\bibinfo {organization} {IEEE},\ \bibinfo {year} {2016})\ pp.\ \bibinfo
  {pages} {1--2}\BibitemShut {NoStop}%
\bibitem [{\citenamefont {Tasker}\ \emph {et~al.}(2023)\citenamefont {Tasker},
  \citenamefont {Frazer}, \citenamefont {Ferranti},\ and\ \citenamefont
  {Matthews}}]{TaskerFrazer23}%
  \BibitemOpen
  \bibfield  {author} {\bibinfo {author} {\bibfnamefont {J.~F.}\ \bibnamefont
  {Tasker}}, \bibinfo {author} {\bibfnamefont {J.}~\bibnamefont {Frazer}},
  \bibinfo {author} {\bibfnamefont {G.}~\bibnamefont {Ferranti}},\ and\
  \bibinfo {author} {\bibfnamefont {J.~C.~F.}\ \bibnamefont {Matthews}},\
  }\bibfield  {title} {\bibinfo {title} {A bi-cmos electronic-photonic
  integrated circuit quantum light detector},\ }\href@noop {} {\bibfield
  {journal} {\bibinfo  {journal} {arXiv:2305.08990}\ } (\bibinfo {year}
  {2023})}\BibitemShut {NoStop}%
\bibitem [{\citenamefont {Paesani}\ \emph {et~al.}(2019)\citenamefont
  {Paesani}, \citenamefont {Borghi}, \citenamefont {Signorini}, \citenamefont
  {Ma\"{\i}nos}, \citenamefont {Pavesi},\ and\ \citenamefont
  {Laing}}]{Paesani19NCOM}%
  \BibitemOpen
  \bibfield  {author} {\bibinfo {author} {\bibfnamefont {S.}~\bibnamefont
  {Paesani}}, \bibinfo {author} {\bibfnamefont {M.}~\bibnamefont {Borghi}},
  \bibinfo {author} {\bibfnamefont {S.}~\bibnamefont {Signorini}}, \bibinfo
  {author} {\bibfnamefont {A.}~\bibnamefont {Ma\"{\i}nos}}, \bibinfo {author}
  {\bibfnamefont {L.}~\bibnamefont {Pavesi}},\ and\ \bibinfo {author}
  {\bibfnamefont {A.}~\bibnamefont {Laing}},\ }\bibfield  {title} {\bibinfo
  {title} {{Near-ideal spontaneous photon sources in silicon quantum
  photonics}},\ }\href@noop {} {\bibfield  {journal} {\bibinfo  {journal}
  {Nature communications}\ }\textbf {\bibinfo {volume} {11}},\ \bibinfo {pages}
  {1} (\bibinfo {year} {2019})}\BibitemShut {NoStop}%
\bibitem [{\citenamefont {Harris}\ \emph {et~al.}(2014)\citenamefont {Harris},
  \citenamefont {Ma}, \citenamefont {Mower}, \citenamefont {Baehr-Jones},
  \citenamefont {Englund}, \citenamefont {Hochberg},\ and\ \citenamefont
  {Galland}}]{Harris2014May}%
  \BibitemOpen
  \bibfield  {author} {\bibinfo {author} {\bibfnamefont {N.~C.}\ \bibnamefont
  {Harris}}, \bibinfo {author} {\bibfnamefont {Y.}~\bibnamefont {Ma}}, \bibinfo
  {author} {\bibfnamefont {J.}~\bibnamefont {Mower}}, \bibinfo {author}
  {\bibfnamefont {T.}~\bibnamefont {Baehr-Jones}}, \bibinfo {author}
  {\bibfnamefont {D.}~\bibnamefont {Englund}}, \bibinfo {author} {\bibfnamefont
  {M.}~\bibnamefont {Hochberg}},\ and\ \bibinfo {author} {\bibfnamefont
  {C.}~\bibnamefont {Galland}},\ }\bibfield  {title} {\bibinfo {title}
  {{Efficient, compact and low loss thermo-optic phase shifter in silicon}},\
  }\href {https://doi.org/10.1364/OE.22.010487} {\bibfield  {journal} {\bibinfo
   {journal} {Opt. Express}\ }\textbf {\bibinfo {volume} {22}},\ \bibinfo
  {pages} {10487} (\bibinfo {year} {2014})}\BibitemShut {NoStop}%
\bibitem [{\citenamefont {Jacques}\ \emph {et~al.}(2019)\citenamefont
  {Jacques}, \citenamefont {Samani}, \citenamefont {El-Fiky}, \citenamefont
  {Patel}, \citenamefont {Xing},\ and\ \citenamefont {Plant}}]{Jacques2019Apr}%
  \BibitemOpen
  \bibfield  {author} {\bibinfo {author} {\bibfnamefont {M.}~\bibnamefont
  {Jacques}}, \bibinfo {author} {\bibfnamefont {A.}~\bibnamefont {Samani}},
  \bibinfo {author} {\bibfnamefont {E.}~\bibnamefont {El-Fiky}}, \bibinfo
  {author} {\bibfnamefont {D.}~\bibnamefont {Patel}}, \bibinfo {author}
  {\bibfnamefont {Z.}~\bibnamefont {Xing}},\ and\ \bibinfo {author}
  {\bibfnamefont {D.~V.}\ \bibnamefont {Plant}},\ }\bibfield  {title} {\bibinfo
  {title} {{Optimization of thermo-optic phase-shifter design and mitigation of
  thermal crosstalk on the SOI platform}},\ }\href
  {https://doi.org/10.1364/OE.27.010456} {\bibfield  {journal} {\bibinfo
  {journal} {Opt. Express}\ }\textbf {\bibinfo {volume} {27}},\ \bibinfo
  {pages} {10456} (\bibinfo {year} {2019})}\BibitemShut {NoStop}%
\bibitem [{\citenamefont {Chakraborty}\ \emph {et~al.}(2020)\citenamefont
  {Chakraborty}, \citenamefont {Carolan}, \citenamefont {Clark}, \citenamefont
  {Bunandar}, \citenamefont {Gilbert}, \citenamefont {Notaros}, \citenamefont
  {Notaros}, \citenamefont {Watts}, \citenamefont {Englund},\ and\
  \citenamefont {Englund}}]{Chakraborty2020Oct}%
  \BibitemOpen
  \bibfield  {author} {\bibinfo {author} {\bibfnamefont {U.}~\bibnamefont
  {Chakraborty}}, \bibinfo {author} {\bibfnamefont {J.}~\bibnamefont
  {Carolan}}, \bibinfo {author} {\bibfnamefont {G.}~\bibnamefont {Clark}},
  \bibinfo {author} {\bibfnamefont {D.}~\bibnamefont {Bunandar}}, \bibinfo
  {author} {\bibfnamefont {G.}~\bibnamefont {Gilbert}}, \bibinfo {author}
  {\bibfnamefont {J.}~\bibnamefont {Notaros}}, \bibinfo {author} {\bibfnamefont
  {J.}~\bibnamefont {Notaros}}, \bibinfo {author} {\bibfnamefont {M.~R.}\
  \bibnamefont {Watts}}, \bibinfo {author} {\bibfnamefont {D.~R.}\ \bibnamefont
  {Englund}},\ and\ \bibinfo {author} {\bibfnamefont {D.~R.}\ \bibnamefont
  {Englund}},\ }\bibfield  {title} {\bibinfo {title} {{Cryogenic operation of
  silicon photonic modulators based on the DC Kerr effect}},\ }\href
  {https://doi.org/10.1364/OPTICA.403178} {\bibfield  {journal} {\bibinfo
  {journal} {Optica}\ }\textbf {\bibinfo {volume} {7}},\ \bibinfo {pages}
  {1385} (\bibinfo {year} {2020})}\BibitemShut {NoStop}%
\bibitem [{\citenamefont {Kaneda}\ \emph {et~al.}(2015)\citenamefont {Kaneda},
  \citenamefont {Christensen}, \citenamefont {Wong}, \citenamefont {Park},
  \citenamefont {McCusker},\ and\ \citenamefont {Kwiat}}]{Kaneda15}%
  \BibitemOpen
  \bibfield  {author} {\bibinfo {author} {\bibfnamefont {F.}~\bibnamefont
  {Kaneda}}, \bibinfo {author} {\bibfnamefont {B.~G.}\ \bibnamefont
  {Christensen}}, \bibinfo {author} {\bibfnamefont {J.~J.}\ \bibnamefont
  {Wong}}, \bibinfo {author} {\bibfnamefont {H.~S.}\ \bibnamefont {Park}},
  \bibinfo {author} {\bibfnamefont {K.~T.}\ \bibnamefont {McCusker}},\ and\
  \bibinfo {author} {\bibfnamefont {P.~G.}\ \bibnamefont {Kwiat}},\ }\bibfield
  {title} {\bibinfo {title} {{Time-multiplexed heralded single-photon
  source}},\ }\href@noop {} {\bibfield  {journal} {\bibinfo  {journal}
  {Optica}\ }\textbf {\bibinfo {volume} {2}},\ \bibinfo {pages} {1010}
  (\bibinfo {year} {2015})}\BibitemShut {NoStop}%
\bibitem [{\citenamefont {Prevedel}\ \emph {et~al.}(2007)\citenamefont
  {Prevedel}, \citenamefont {Walther}, \citenamefont {Tiefenbacher},
  \citenamefont {B{\"o}hi}, \citenamefont {Kaltenbaek}, \citenamefont
  {Jennewein},\ and\ \citenamefont {Zeilinger}}]{prevedel2007high}%
  \BibitemOpen
  \bibfield  {author} {\bibinfo {author} {\bibfnamefont {R.}~\bibnamefont
  {Prevedel}}, \bibinfo {author} {\bibfnamefont {P.}~\bibnamefont {Walther}},
  \bibinfo {author} {\bibfnamefont {F.}~\bibnamefont {Tiefenbacher}}, \bibinfo
  {author} {\bibfnamefont {P.}~\bibnamefont {B{\"o}hi}}, \bibinfo {author}
  {\bibfnamefont {R.}~\bibnamefont {Kaltenbaek}}, \bibinfo {author}
  {\bibfnamefont {T.}~\bibnamefont {Jennewein}},\ and\ \bibinfo {author}
  {\bibfnamefont {A.}~\bibnamefont {Zeilinger}},\ }\bibfield  {title} {\bibinfo
  {title} {High-speed linear optics quantum computing using active
  feed-forward},\ }\href@noop {} {\bibfield  {journal} {\bibinfo  {journal}
  {Nature}\ }\textbf {\bibinfo {volume} {445}},\ \bibinfo {pages} {65}
  (\bibinfo {year} {2007})}\BibitemShut {NoStop}%
\bibitem [{\citenamefont {Wang}\ \emph {et~al.}(2017)\citenamefont {Wang},
  \citenamefont {Paesani}, \citenamefont {Santagati}, \citenamefont {Knauer},
  \citenamefont {Gentile}, \citenamefont {Wiebe}, \citenamefont {Petruzzella},
  \citenamefont {O’Brien}, \citenamefont {Rarity}, \citenamefont {Laing}
  \emph {et~al.}}]{wang2017experimental}%
  \BibitemOpen
  \bibfield  {author} {\bibinfo {author} {\bibfnamefont {J.}~\bibnamefont
  {Wang}}, \bibinfo {author} {\bibfnamefont {S.}~\bibnamefont {Paesani}},
  \bibinfo {author} {\bibfnamefont {R.}~\bibnamefont {Santagati}}, \bibinfo
  {author} {\bibfnamefont {S.}~\bibnamefont {Knauer}}, \bibinfo {author}
  {\bibfnamefont {A.~A.}\ \bibnamefont {Gentile}}, \bibinfo {author}
  {\bibfnamefont {N.}~\bibnamefont {Wiebe}}, \bibinfo {author} {\bibfnamefont
  {M.}~\bibnamefont {Petruzzella}}, \bibinfo {author} {\bibfnamefont {J.~L.}\
  \bibnamefont {O’Brien}}, \bibinfo {author} {\bibfnamefont {J.~G.}\
  \bibnamefont {Rarity}}, \bibinfo {author} {\bibfnamefont {A.}~\bibnamefont
  {Laing}}, \emph {et~al.},\ }\bibfield  {title} {\bibinfo {title}
  {Experimental quantum hamiltonian learning},\ }\href@noop {} {\bibfield
  {journal} {\bibinfo  {journal} {Nature Physics}\ }\textbf {\bibinfo {volume}
  {13}},\ \bibinfo {pages} {551} (\bibinfo {year} {2017})}\BibitemShut
  {NoStop}%
\bibitem [{\citenamefont {Reed}\ \emph {et~al.}(2010)\citenamefont {Reed},
  \citenamefont {Mashanovich}, \citenamefont {Gardes},\ and\ \citenamefont
  {Thomson}}]{Reed10}%
  \BibitemOpen
  \bibfield  {author} {\bibinfo {author} {\bibfnamefont {G.~T.}\ \bibnamefont
  {Reed}}, \bibinfo {author} {\bibfnamefont {G.}~\bibnamefont {Mashanovich}},
  \bibinfo {author} {\bibfnamefont {F.}~\bibnamefont {Gardes}},\ and\ \bibinfo
  {author} {\bibfnamefont {D.}~\bibnamefont {Thomson}},\ }\bibfield  {title}
  {\bibinfo {title} {{Silicon optical modulators}},\ }\href@noop {} {\bibfield
  {journal} {\bibinfo  {journal} {Nature Photonics}\ }\textbf {\bibinfo
  {volume} {4}},\ \bibinfo {pages} {518} (\bibinfo {year} {2010})}\BibitemShut
  {NoStop}%
\bibitem [{\citenamefont {Reed}\ and\ \citenamefont
  {Jason~Png}(2005)}]{Reed2005Jan}%
  \BibitemOpen
  \bibfield  {author} {\bibinfo {author} {\bibfnamefont {G.~T.}\ \bibnamefont
  {Reed}}\ and\ \bibinfo {author} {\bibfnamefont {C.~E.}\ \bibnamefont
  {Jason~Png}},\ }\bibfield  {title} {\bibinfo {title} {{Silicon optical
  modulators}},\ }\href {https://doi.org/10.1016/S1369-7021(04)00678-9}
  {\bibfield  {journal} {\bibinfo  {journal} {Mater. Today}\ }\textbf {\bibinfo
  {volume} {8}},\ \bibinfo {pages} {40} (\bibinfo {year} {2005})}\BibitemShut
  {NoStop}%
\bibitem [{\citenamefont {Reed}\ \emph {et~al.}(2014)\citenamefont {Reed},
  \citenamefont {Mashanovich}, \citenamefont {Gardes}, \citenamefont
  {Nedeljkovic}, \citenamefont {Hu}, \citenamefont {Thomson}, \citenamefont
  {Li}, \citenamefont {Wilson}, \citenamefont {Chen},\ and\ \citenamefont
  {Hsu}}]{Reed2014Aug}%
  \BibitemOpen
  \bibfield  {author} {\bibinfo {author} {\bibfnamefont {G.~T.}\ \bibnamefont
  {Reed}}, \bibinfo {author} {\bibfnamefont {G.~Z.}\ \bibnamefont
  {Mashanovich}}, \bibinfo {author} {\bibfnamefont {F.~Y.}\ \bibnamefont
  {Gardes}}, \bibinfo {author} {\bibfnamefont {M.}~\bibnamefont {Nedeljkovic}},
  \bibinfo {author} {\bibfnamefont {Y.}~\bibnamefont {Hu}}, \bibinfo {author}
  {\bibfnamefont {D.~J.}\ \bibnamefont {Thomson}}, \bibinfo {author}
  {\bibfnamefont {K.}~\bibnamefont {Li}}, \bibinfo {author} {\bibfnamefont
  {P.~R.}\ \bibnamefont {Wilson}}, \bibinfo {author} {\bibfnamefont {S.-W.}\
  \bibnamefont {Chen}},\ and\ \bibinfo {author} {\bibfnamefont {S.~S.}\
  \bibnamefont {Hsu}},\ }\bibfield  {title} {\bibinfo {title} {{Recent
  breakthroughs in carrier depletion based silicon optical modulators}},\
  }\href {https://doi.org/10.1515/nanoph-2013-0016} {\bibfield  {journal}
  {\bibinfo  {journal} {Nanophotonics}\ }\textbf {\bibinfo {volume} {3}},\
  \bibinfo {pages} {229} (\bibinfo {year} {2014})}\BibitemShut {NoStop}%
\bibitem [{\citenamefont {Sibson}\ \emph {et~al.}(2017)\citenamefont {Sibson},
  \citenamefont {Kennard}, \citenamefont {Stanisic}, \citenamefont {Erven},
  \citenamefont {O{'}Brien},\ and\ \citenamefont {Thompson}}]{Sibson2017Feb}%
  \BibitemOpen
  \bibfield  {author} {\bibinfo {author} {\bibfnamefont {P.}~\bibnamefont
  {Sibson}}, \bibinfo {author} {\bibfnamefont {J.~E.}\ \bibnamefont {Kennard}},
  \bibinfo {author} {\bibfnamefont {S.}~\bibnamefont {Stanisic}}, \bibinfo
  {author} {\bibfnamefont {C.}~\bibnamefont {Erven}}, \bibinfo {author}
  {\bibfnamefont {J.~L.}\ \bibnamefont {O{'}Brien}},\ and\ \bibinfo {author}
  {\bibfnamefont {M.~G.}\ \bibnamefont {Thompson}},\ }\bibfield  {title}
  {\bibinfo {title} {{Integrated silicon photonics for high-speed quantum key
  distribution}},\ }\href {https://doi.org/10.1364/OPTICA.4.000172} {\bibfield
  {journal} {\bibinfo  {journal} {Optica}\ }\textbf {\bibinfo {volume} {4}},\
  \bibinfo {pages} {172} (\bibinfo {year} {2017})}\BibitemShut {NoStop}%
\bibitem [{\citenamefont {Rahim}\ \emph {et~al.}(2019)\citenamefont {Rahim},
  \citenamefont {Goyvaerts}, \citenamefont {Szelag}, \citenamefont {Fedeli},
  \citenamefont {Absil}, \citenamefont {Aalto}, \citenamefont {Harjanne},
  \citenamefont {Littlejohns}, \citenamefont {Reed}, \citenamefont {Winzer},
  \citenamefont {Lischke}, \citenamefont {Zimmermann}, \citenamefont {Knoll},
  \citenamefont {Geuzebroek}, \citenamefont {Leinse}, \citenamefont
  {Geiselmann}, \citenamefont {Zervas}, \citenamefont {Jans}, \citenamefont
  {Stassen}, \citenamefont {Dom{\ifmmode\acute{\imath}\else\'{\i}\fi}nguez},
  \citenamefont {Mu{\ifmmode\tilde{n}\else\~{n}\fi}oz}, \citenamefont
  {Domenech}, \citenamefont {Giesecke}, \citenamefont {Lemme},\ and\
  \citenamefont {Baets}}]{Rahim2019May}%
  \BibitemOpen
  \bibfield  {author} {\bibinfo {author} {\bibfnamefont {A.}~\bibnamefont
  {Rahim}}, \bibinfo {author} {\bibfnamefont {J.}~\bibnamefont {Goyvaerts}},
  \bibinfo {author} {\bibfnamefont {B.}~\bibnamefont {Szelag}}, \bibinfo
  {author} {\bibfnamefont {J.-M.}\ \bibnamefont {Fedeli}}, \bibinfo {author}
  {\bibfnamefont {P.}~\bibnamefont {Absil}}, \bibinfo {author} {\bibfnamefont
  {T.}~\bibnamefont {Aalto}}, \bibinfo {author} {\bibfnamefont
  {M.}~\bibnamefont {Harjanne}}, \bibinfo {author} {\bibfnamefont
  {C.}~\bibnamefont {Littlejohns}}, \bibinfo {author} {\bibfnamefont
  {G.}~\bibnamefont {Reed}}, \bibinfo {author} {\bibfnamefont {G.}~\bibnamefont
  {Winzer}}, \bibinfo {author} {\bibfnamefont {S.}~\bibnamefont {Lischke}},
  \bibinfo {author} {\bibfnamefont {L.}~\bibnamefont {Zimmermann}}, \bibinfo
  {author} {\bibfnamefont {D.}~\bibnamefont {Knoll}}, \bibinfo {author}
  {\bibfnamefont {D.}~\bibnamefont {Geuzebroek}}, \bibinfo {author}
  {\bibfnamefont {A.}~\bibnamefont {Leinse}}, \bibinfo {author} {\bibfnamefont
  {M.}~\bibnamefont {Geiselmann}}, \bibinfo {author} {\bibfnamefont
  {M.}~\bibnamefont {Zervas}}, \bibinfo {author} {\bibfnamefont
  {H.}~\bibnamefont {Jans}}, \bibinfo {author} {\bibfnamefont {A.}~\bibnamefont
  {Stassen}}, \bibinfo {author} {\bibfnamefont {C.}~\bibnamefont
  {Dom{\ifmmode\acute{\imath}\else\'{\i}\fi}nguez}}, \bibinfo {author}
  {\bibfnamefont {P.}~\bibnamefont {Mu{\ifmmode\tilde{n}\else\~{n}\fi}oz}},
  \bibinfo {author} {\bibfnamefont {D.}~\bibnamefont {Domenech}}, \bibinfo
  {author} {\bibfnamefont {A.~L.}\ \bibnamefont {Giesecke}}, \bibinfo {author}
  {\bibfnamefont {M.~C.}\ \bibnamefont {Lemme}},\ and\ \bibinfo {author}
  {\bibfnamefont {R.}~\bibnamefont {Baets}},\ }\bibfield  {title} {\bibinfo
  {title} {{Open-Access Silicon Photonics Platforms in Europe}},\ }\href
  {https://doi.org/10.1109/JSTQE.2019.2915949} {\bibfield  {journal} {\bibinfo
  {journal} {IEEE J. Sel. Top. Quantum Electron.}\ }\textbf {\bibinfo {volume}
  {25}},\ \bibinfo {pages} {1} (\bibinfo {year} {2019})}\BibitemShut {NoStop}%
\bibitem [{\citenamefont {Sharping}\ \emph {et~al.}(2006)\citenamefont
  {Sharping}, \citenamefont {Lee}, \citenamefont {Foster}, \citenamefont
  {Turner}, \citenamefont {Schmidt}, \citenamefont {Lipson}, \citenamefont
  {Gaeta},\ and\ \citenamefont {Kumar}}]{Sharping2006Dec}%
  \BibitemOpen
  \bibfield  {author} {\bibinfo {author} {\bibfnamefont {J.~E.}\ \bibnamefont
  {Sharping}}, \bibinfo {author} {\bibfnamefont {K.~F.}\ \bibnamefont {Lee}},
  \bibinfo {author} {\bibfnamefont {M.~A.}\ \bibnamefont {Foster}}, \bibinfo
  {author} {\bibfnamefont {A.~C.}\ \bibnamefont {Turner}}, \bibinfo {author}
  {\bibfnamefont {B.~S.}\ \bibnamefont {Schmidt}}, \bibinfo {author}
  {\bibfnamefont {M.}~\bibnamefont {Lipson}}, \bibinfo {author} {\bibfnamefont
  {A.~L.}\ \bibnamefont {Gaeta}},\ and\ \bibinfo {author} {\bibfnamefont
  {P.}~\bibnamefont {Kumar}},\ }\bibfield  {title} {\bibinfo {title}
  {{Generation of correlated photons in nanoscale silicon waveguides}},\ }\href
  {https://doi.org/10.1364/OE.14.012388} {\bibfield  {journal} {\bibinfo
  {journal} {Opt. Express}\ }\textbf {\bibinfo {volume} {14}},\ \bibinfo
  {pages} {12388} (\bibinfo {year} {2006})}\BibitemShut {NoStop}%
\bibitem [{\citenamefont {Rosenfeld}\ \emph {et~al.}(2020)\citenamefont
  {Rosenfeld}, \citenamefont {Sulway}, \citenamefont {Sinclair}, \citenamefont
  {Anant}, \citenamefont {Thompson}, \citenamefont {Rarity},\ and\
  \citenamefont {Silverstone}}]{Rosenfeld20}%
  \BibitemOpen
  \bibfield  {author} {\bibinfo {author} {\bibfnamefont {L.~M.}\ \bibnamefont
  {Rosenfeld}}, \bibinfo {author} {\bibfnamefont {D.~A.}\ \bibnamefont
  {Sulway}}, \bibinfo {author} {\bibfnamefont {G.~F.}\ \bibnamefont
  {Sinclair}}, \bibinfo {author} {\bibfnamefont {V.}~\bibnamefont {Anant}},
  \bibinfo {author} {\bibfnamefont {M.~G.}\ \bibnamefont {Thompson}}, \bibinfo
  {author} {\bibfnamefont {J.~G.}\ \bibnamefont {Rarity}},\ and\ \bibinfo
  {author} {\bibfnamefont {J.~W.}\ \bibnamefont {Silverstone}},\ }\bibfield
  {title} {\bibinfo {title} {{Mid-infrared quantum optics in silicon}},\
  }\href@noop {} {\bibfield  {journal} {\bibinfo  {journal} {Optics Express}\
  }\textbf {\bibinfo {volume} {28}},\ \bibinfo {pages} {37092} (\bibinfo {year}
  {2020})}\BibitemShut {NoStop}%
\bibitem [{\citenamefont {Rudolph}\ \emph {et~al.}(2018)\citenamefont
  {Rudolph}, \citenamefont {Thompson}, \citenamefont {Matthews},\ and\
  \citenamefont {Bonneau}}]{rudolph2018optical}%
  \BibitemOpen
  \bibfield  {author} {\bibinfo {author} {\bibfnamefont {T.}~\bibnamefont
  {Rudolph}}, \bibinfo {author} {\bibfnamefont {M.}~\bibnamefont {Thompson}},
  \bibinfo {author} {\bibfnamefont {J.}~\bibnamefont {Matthews}},\ and\
  \bibinfo {author} {\bibfnamefont {D.}~\bibnamefont {Bonneau}},\ }\href@noop
  {} {\bibinfo {title} {Optical apparatus and method for outputting one or more
  photons}} (\bibinfo {year} {2018}),\ \bibinfo {note} {uS Patent
  9,952,482}\BibitemShut {NoStop}%
\end{thebibliography}
\end{document}